\documentstyle[prl,aps]{revtex}
\begin{document}
\draft
\title{Optical Solitary Waves in the Higher Order Nonlinear Schr\"{o}dinger
Equation}
\author{M. Gedalin, T.C. Scott, and Y.B. Band}
\address{Departments of Chemistry and Physics,
Ben-Gurion University of the Negev,
\\84105 Beer-Sheva, Israel}
\maketitle
\begin{abstract}
We study solitary wave solutions of the higher order nonlinear
Schr\"{o}dinger equation for the propagation of short light pulses in
an optical fiber.  Using a scaling transformation we reduce the
equation to a two-parameter canonical form.  Solitary wave ($1$-soliton) 
solutions {\it always} exist provided easily met inequality constraints on 
the parameters in the equation are satisfied.  Conditions for the existence
of $N$-soliton solutions ($N\ge 2$) are determined; when these
conditions are met the equation becomes the modified KdV equation.  A
proper subset of these conditions meet the Painlev\'{e} plausibility
conditions for integrability.
\end{abstract}
\pacs{42.81.Dp, 02.30.Jr, 42.65.Tg, 42.79.Sz}

The propagation of nonlinear waves in dispersive media is of great 
interest since nonlinear dispersive systems are ubiquitous in nature.
Propagation of ultrashort light pulses in optical fibers is of particular 
interest because of the common expectation that solitary waves may be 
of extensive use in telecommunication and even revolutionarize it.  The
existence of solitary wave solutions implies perfect balance between
nonlinearity and dispersion which usually requires rather specific
conditions and cannot be established in general. The objective of the
present paper is to study the conditions under which the existence of
solitary waves is guaranteed for ultrashort pulses.

The propagation of light pulses in fibers is well described by the
higher order nonlinear Schr\"{o}dinger equation (HONSE) 
\cite{Kodama,Mitschke,Hasegawa,Agrawal}, a partial differential equation
(PDE) whose right hand side includes the effects of group velocity 
dispersion, self-phase modulation, third order dispersion, self-steepening, 
and self-frequency shifting via stimulated Raman scattering, respectively:
\begin{equation}
E_z = i(\alpha_1 E_{tt} + \alpha_2 |E|^2 E) + \alpha_3 E_{ttt} \
+ \alpha_4 (|E|^2 E)_t + \alpha_5 (|E|^2)_t E. \label{honse}
\end{equation}
When the last three terms are omitted this propagation equation for the
slowly varying envelope of the electric field, $E$, reduces to the
nonlinear Schr\"{o}dinger equations (NSE), which is integrable (meaning
it not only admits $N$-solitary wave solutions, but that the evolution of any
initial condition is known in principle) \cite{Agrawal,AS81,Hasegawa2}.  
We call these $N$-solitary wave solutions $N$-solitons, and mean by this
that the solitary waves scatter elastically and asymptotically preserve 
their shape upon undergoing collisions, just like true solitons.
However, for short duration pulses the last three terms are
non-negligible and should be retained. In general, the presence or
absence of solitary wave solutions depends on the coefficients $\alpha$
appearing in Eq.~(\ref{honse}), and therefore, on the specific
nonlinear and dispersive features of the medium.  
Here, we reduce the HONSE to a two-parameter equation and derive a general 
solitary wave ($1$-soliton) solution.  We determine conditions when 
$N$-soliton solutions exist and display the solutions.
We also study the Painlev\'{e} plausibility conditions for
integrability and show that these are only a proper subset of the 
conditions necessary for $N$-soliton solutions to exist.

We begin by scaling the HONSE, letting $E = b_1 A$, $z = b_2 \zeta$, and
$t = b_3 \tau$.  Substituting into the HONSE we obtain
\begin{equation}
A_\zeta = i((b_2 \alpha_1/b_3^2) A_{\tau\tau} + (b_1^2 b_2) 
\alpha_2 |A|^2 A)+ 
(b_2\alpha_3/b_3^3) A_{\tau\tau\tau} 
+ (b_1^2 b_2 \alpha_4/b_3) (|A|^2 A)_\tau
 + (b_1^2 b_2 \alpha_5/b_3) (|A|^2)_\tau A. \label{honse1}
\end{equation}
Choosing $b_1=(\alpha_1^3/(\alpha_2\alpha_3^2))^{1/2}$, 
$b_2=\alpha_3^2/\alpha_1^3$, and $b_3 = \alpha_3/\alpha_1$,
we can set the coefficients of the first, second and third terms on the
right hand side of Eq.~(\ref{honse1}) to unity, so that the HONSE becomes
\begin{equation} 
A_\zeta = i(A_{\tau\tau} + |A|^2 A) + A_{\tau\tau\tau}
+ \gamma_1 (|A|^2 A)_\tau + \gamma_2 (|A|^2)_\tau A, \label{honse2}
\end{equation}
where $\gamma_1 = b_1^2b_2\alpha_4/b_3=\alpha_4\alpha_1/\alpha_2 
\alpha_3$ and $\gamma_2 = b_1^2 b_2
\alpha_5/b_3=\alpha_5\alpha_1/\alpha_2 
\alpha_3$.

A solution to Eq.~(\ref{honse2}) of the form
$A(\zeta,\tau)=y(\tau+\beta\zeta) \exp[i(\kappa \zeta - \Omega 
\tau)]$, with $y$ real,  
exists regardless of the values of the parameters $\gamma_1$ and
$\gamma_2$ as can be easily demonstrated by substituting this form into
Eq.~(\ref{honse2}) and equating the real 
and imaginary parts of the resulting
equation.  This procedure yields the following necessary and sufficient
condition on $\Omega$, and an equation for $\kappa$ in terms of $\Omega$:
\begin{eqnarray}
&& \Omega=\frac{3\gamma_1+2\gamma_2-3}{6(\gamma_1+\gamma_2)}, \qquad
(\Omega\not=1/3), \label{first}\\
&& \kappa-\Omega^3+\Omega^2 = (\beta + 3\Omega^2 - 2\Omega)(1-3\Omega).
\label{second}
\end{eqnarray}
The function $y(\xi)$ ($\xi = \tau+\beta\zeta$) satisfies the ordinary 
differential equation
\begin{equation}
y_{\xi\xi} = (\beta + 3\Omega^2 - 2\Omega) y
-\frac{3\gamma_1+2\gamma_2}{3} y^3 ,
\label{ode}\end{equation}
whose solution is given generally in terms of doubly periodic elliptic 
functions. For zero energy (i.e., for \\ $y_{\xi}^2-(\beta + 
3\Omega^2 - 2\Omega)y^2+(3\gamma_1+2\gamma_2)y^4/6=0$) we find 
the solitary wave solution
\begin{eqnarray}
&&A(\zeta,\tau) = \left(\frac{6(\beta + 3\Omega^2 
- 2\Omega)}{3\gamma_1+2\gamma_2}\right)^{1/2}
\cosh^{-1}[(\beta + 3\Omega^2 - 2\Omega)^{1/2}(\tau+\beta\zeta)] 
\nonumber\\
&& \times \exp\{i[((\beta + 3\Omega^2 
- 2\Omega)(1-3\Omega)+\Omega^3-\Omega^2)
 \zeta - \Omega \tau]\},
\end{eqnarray}
provided $(\beta + 3\Omega^2 - 2\Omega)>0$ and $3\gamma_1+2\gamma_2>0$.
Thus, solitary wave solutions always exist 
(in contrast with what is implied in
Ref.~\onlinecite{Porsezian}) provided $3\gamma_1+2\gamma_2>0$.
For $\Omega=0$ (i.e., for $3\gamma_1+2\gamma_2 = 3$) the solitary wave
solution reduces to 
$A(\zeta,\tau)=2^{1/2}\eta \cosh^{-1}(\eta(\tau+\eta^2\zeta)) 
\exp(i\eta^2\zeta)$ where $\kappa=\beta \equiv \eta^2>0$.

The case of $\Omega=1/3$ is very special, not only because 
Eq.~(\ref{first})
(which should be written in the form
$3\gamma_1+2\gamma_2=3(1-\gamma_1\Omega)/(1-3\Omega)$) is not
applicable, but also because Eq.~(\ref{honse2}) is expected to be 
integrable for this case, as we
shall show below using a Painlev\'{e} analysis \cite{Weiss,Musette,Ablowitz}.
The Painlev\'{e} condition for integrability (see below)
\begin{equation}
\gamma_1 = 3, \gamma_2 = -3/2,   \label{int}
\end{equation}
(which differs from the result claimed in 
Ref.~\onlinecite{Porsezian}) yields 
$\Omega = 1/3$ and $\kappa = -2/27$, and the solitary wave solution takes 
the form
\begin{equation}
A(\zeta,\tau) = (\beta-1/3)^{1/2}
\cosh^{-1}[(\beta-1/3)^{1/2}(\tau+\beta\zeta)] \exp[-i(2\zeta/27 + 
\tau/3)].
\end{equation}

It is of interest to compare the solitary wave solutions for different values
of parameters $\gamma_1$ and $\gamma_2$.  All solitary waves have
intensity profiles of the form $I(\zeta,\tau) =  |A(\zeta,\tau)|^2 = I_s
\cosh^{-2}[(\tau+\beta\zeta)/\tau_s]$.  
The solitary wave width $\tau_s$ and its 
intensity $I_s$ are related by the expression 
$I_s \tau_s^2 = 6/(3\gamma_1+2\gamma_2)$ (where $3\gamma_1+2\gamma_2 > 0$ 
for a solitary wave solution).  For
comparison, recall that $I_s \tau_s^2 = 2$ for the NSE solitary wave
($\alpha_3=\alpha_4=\alpha_5=0$), so that for equal $\tau_s$, 
$I_s^{\mbox{HONSE}}/I_s^{\mbox{NSE}}
=3/(3\gamma_1+ 2\gamma_2)$.  The width 
$\tau_s$ and $\beta$ (the negative of 
the inverse-velocity in the coordinate 
system moving with the group velocity of the light pulse, i.e., the
solitary wave velocity $v_s=v_g-\beta^{-1}$ 
where $v_g$ is the group velocity) are 
related by the equation $\tau_s = (\beta + 3\Omega^2 - 2\Omega)^{-1/2}$ 
where $\Omega$ is given by Eq.~(\ref{first}).

As is well-known, existence of solitary wave solutions does not guarantee 
existence of $N$-soliton solutions with $N>1$. In the absence of 
the inverse scattering solution for the HONSE, the Hirota method 
\cite{Hirota}, based on the cutoff of the Pad\'{e}-approximation 
\cite{Ba70} $A=G/F$, $F$ being real, is often useful. Direct substitution 
of this representation into the PDE in Eq.~(\ref{honse2}) can be expedited 
using Hirota $D$ operators \cite{Hirota} (defined by its operation on 
bilinear forms
$D_t(f\cdot g):=
(\partial/\partial t - \partial/\partial t')f(t)g(t')|_{t=t'}$)
and by separation of the linear part of the PDE to yield:
\begin{eqnarray}
&& (D_\zeta-iD_\tau^2-D_\tau^3) (G\cdot F)=0, \label{hirota1}\\
&& (G\cdot F)[-iD_\tau^2(F\cdot F) + i (G^*\cdot G) + (\gamma_1+\gamma_2) 
D_\tau(G^*\cdot G)] \nonumber\\
&& +D_\tau(G\cdot F)[-3D_\tau^2(F\cdot F)
+(3\gamma_1+2\gamma_2)(G^*\cdot G)]=0. 
\label{hirota2}
\end{eqnarray}
The standard algorithm is to further substitute a 2-soliton solution in 
a power series in $\epsilon$ of the form
\begin{eqnarray}
&& G=\epsilon(\exp(\theta_1) +\exp(\theta_2)) +\epsilon^2 G^{(2)} + 
\ldots, \label{g}\\
&& F=1 +\epsilon F^{(1)} + \epsilon^2 F^{(2)} + \ldots, \label{f}\\
&& \theta_i=p_i\zeta+q_i\tau+\phi_i, \quad i=1,2, \label{theta}
\end{eqnarray}
and require that the series truncate.
 
The Hirota method is well adapted to the case when 
Eq.~(\ref{hirota2}) reduces to a bilinear equation 
or can be split into two or more independent (and consistent) bilinear 
equations. Unfortunately, further splitting of Eq.~(\ref{hirota2}) into two
bilinear equations is impossible in the general case (the direct 
na\"{i}ve splitting of Eq.~(\ref{hirota2}) into two equations corresponding 
to the two expressions in square brackets \cite{LW94,Porsezian} is 
incorrect, since it gives $p_1=p_2$ and $q_1=q_2$ for the $2$-soliton 
substitution (\ref{g}) which is forbidden). Direct analysis of the 
multilinear equation (\ref{hirota2}) is not successful either. We shall 
try further reduction by substituting 
\begin{equation}
G=\bar{G}\exp [i(\kappa\zeta -\Omega\tau)] \label{gbar}
\end{equation}
into Eqs.~(\ref{hirota1}) and (\ref{hirota2}). Choosing 
$\kappa=\Omega^2-\Omega^3$ we retain the structure of Eq.~(\ref{hirota1}), 
while Eq~(\ref{hirota2}) takes the following form:
\begin{eqnarray}
&& (\bar{G}\cdot F) [-i(1-3\Omega)D_\tau^2(F\cdot F) + i(1 -\Omega\gamma_1) 
(\bar{G}^*\cdot \bar{G}) + (\gamma_1 +\gamma_2) D_\tau (\bar{G}^*\cdot 
\bar{G})]\nonumber\\
&& +D_\tau(\bar{G}\cdot F) [-3D_\tau^2(F\cdot F) +(3 \gamma_1 
+2\gamma_2) (\bar{G}^*\cdot \bar{G})]=0. \label{hirota3}
\end{eqnarray}

Selecting $\Omega=1/3$ or $\Omega=1/\gamma_1$ eliminates  the terms $\propto 
D_\tau^2(F\cdot F)$ or $(\bar{G}^*\cdot \bar{G})$ in the first part of 
Eq.~(\ref{hirota3}), respectively. Further simplification can be 
achieved in the particular case $\gamma_1=3$ when the choice $\Omega=1/3$
leads to the following equation
\begin{equation}
D_\tau(\bar{G}\cdot F)[-3D_\tau^2(F\cdot F)+ (9+2\gamma_2) (\bar{G}^* 
\cdot \bar{G})] 
+(\bar{G}\cdot F) (3 + \gamma_2) D_\tau(\bar{G}^*\cdot \bar{G})=0.
 \label{hirota2a}
\end{equation}
Note that the first Hirota equation (\ref{hirota1}) implies 
that any $N$-soliton solution which takes the Hirota form 
can be written as $G/F=\Lambda(\gamma_2) {\cal G}/F$, where 
${\cal G}$ and $F$ are independent of the parameter $\gamma_2$ (cf. 
\cite{Hirota,LW94,Porsezian}). 
Substituting this into (\ref{hirota2a}) shows that this is possible only if 
$\gamma_2=-3$ or $\ln({\cal G}^*/{\cal G})=\mbox{const}$. 
The first choice corresponds to the 
case described in Ref.~\onlinecite{Hirota}. We shall consider here the 
second choice  which leaves $\gamma_2$ arbitrary. 
 
Returning to Eqs.~(\ref{honse2}) we substitute
\begin{equation}
A={\cal E}(\zeta,\tau)\,\exp[i(\kappa\zeta -\Omega\tau)]. \label{solution}
\end{equation} 
which  gives:
\begin{eqnarray}
&&{\cal E}_\zeta = - i(\kappa+\Omega^2 -\Omega^3) {\cal E} 
+ (2\Omega - 3\Omega^2) {\cal E}_\tau + 
i(1-3\Omega ) {\cal E}_{\tau\tau} \nonumber\\
&&+ i(1-\Omega\gamma_1) |{\cal E}|^2{\cal E} 
+{\cal E}_{\tau\tau\tau} 
+\gamma_1(|{\cal E}|^2{\cal E})_\tau + \gamma_2 {\cal E} 
(|{\cal E}|^2)_\tau.
\end{eqnarray}
Choosing $\kappa=\Omega^3-\Omega^2$ and making the 
coordinate transformation $T=\tau+(2\Omega 
- 3\Omega^2)\zeta$, eliminates the terms 
$\propto {\cal E}$ and ${\cal E}_\tau$. 
Selecting $\Omega=1/3$ or $\Omega=1/\gamma_1$, eliminates terms 
$\propto {\cal E}_{\tau\tau}$ or $|{\cal E}|^2{\cal E}$, respectively. 
In the particular case $\gamma_1=3$,  $\Omega=1/3$, one arrives at the 
following complex modified KdV equation:
\begin{equation}
{\cal E}_\zeta={\cal E}_{TTT} +3(|{\cal E}|^2{\cal E})_T +\gamma_2 
{\cal E}(|{\cal E}|^2)_T, \label{mkdv}
\end{equation}
which is more general than the equation considered in Ref. 
\onlinecite{Sa91} since it is for arbitrary $\gamma_2$.  
It is easy to see that, for the special case $\gamma_1=3$, 
Eq.~(\ref{honse2}) has an $N$-soliton solution of the form:
\begin{equation}
A = B(\zeta,T) \exp[i(-2\zeta/27  -\tau/3+\psi)], \quad 
T=\tau+\zeta/3, \quad \psi=\mbox{const}, \label{nsoliton}
\end{equation}
where $B(\zeta,T)$ is the $N$-soliton solution of the real
modified KdV equation 
\begin{equation}
B_\zeta=B_{TTT}+(9+2\gamma_2) B^2B_T, \label{mkdv1}
\end{equation}
and can be written as \cite{AS81}
\begin{eqnarray}
 && B=i(\frac{6}{9+2\gamma_2})^{1/2}(\ln(f^*/f))_T, \label{n1}\\
 && f=\sum_{\mu=0,1}\exp(\sum_{i=1}^N\mu_i(\eta_i +i\pi/2) +\sum_{1\leq 
 i<j}^N \mu_i \mu_jA_{ij}), \label{n2}\\
&& \eta_i=q_iT+q_i^3\zeta +\eta_i^{(0)}, \quad A_{ij}= \ln(\frac{q_i 
-q_j}{q_i+q_j})^2. \label{n3}
\end{eqnarray} 
This $N$-soliton solution differs from that proposed in 
Ref.~\onlinecite{Porsezian} (which does not fulfill the corresponding 
nonlinear equation). Note that the existence of $N$-soliton solutions 
does not in general imply integrability. However, the integrability of the 
complex modified KdV in the special case of $\gamma_1=3$ and 
$\gamma_2=-3/2$ has been demonstrated in Ref.~\onlinecite{Sa91}.
 
Finally, we study the integrability of the HONSE, applying the 
Painlev\'{e} analysis \cite{Weiss,Musette,Ablowitz}.  It is widely 
believed that possession of the Painlev\'{e} feature is a sufficient 
criterion for integrability (see discussion in 
Refs.~\onlinecite{Ne85,Ra89}).  The PDE in Eq.~(3) can be analyzed to 
ascertain whether it is integrable by seeking a solution of the PDE in 
the Painlev\'{e} form $A(\zeta,\tau) = (a(\zeta,\tau))^{-\sigma} 
\sum_{m=0}^{\infty}b_m(\zeta)(a(\zeta,\tau))^m$, where $b_m(\zeta)$ 
are analytic functions of $\zeta$ in the neighborhood of a 
noncharacteristic movable singularity manifold defined by 
$a(\zeta,\tau) = \tau-f(\zeta)$.  Following Ref.~\onlinecite{Ra89} we 
substitute into Eq.~(3) the following Laurent series
\begin{eqnarray} 
&& A=(\tau-\tau_0)^{-\sigma}\sum_{m=0} 
R_m(\zeta)(\tau-\tau_0)^m, \label{pan1}\\
&& A^*=(\tau-\tau_0)^{-\sigma}\sum_{m=0} 
S_m(\zeta)(\tau-\tau_0)^m, \label{pan2}
\end{eqnarray}
(in the vicinity of a movable singular point $\tau=\tau_0(\zeta)$.  
Inspection of the strongest singularity immediately gives $\sigma=1$ 
and $R_0S_0=-6/(3\gamma_1 +2\gamma_2)$.  The subsequent substitution 
of (\ref{pan1}) and (\ref{pan2}) into Eq.~(3) should allow 
identification of the other $R_m$ and $S_m$ leaving exactly 6 
arbitrary functions (including $\tau_0(z)$) undetermined.  Analysis of 
resonances is carried out by making an auxiliary substitution of the 
form $A = R_0(\tau-\tau_0)^{-1} + R_p(\tau-\tau_0)^{p-1}$ (and 
similarly for $A^*$) into the PDE and retaining only linear terms in 
$R_p$ (cf.  Ref.~\onlinecite{AS81,Porsezian}).  This procedure shows 
that resonances occur at $p=-1,0,3,4$, and $3\pm 
2\sqrt{-\gamma_1/(3\gamma_1 +2\gamma_2)}$.  The resonances at $p=-1$ 
and $p=0$ correspond to the arbitrariness of $\tau_0$ and $S_0/R_0$, 
respectively.  Requiring that the resonant indices $p$ be integers 
\cite{Ra89} one finds that either $\gamma_1=-2\gamma_2$ or 
$\gamma_1=-\gamma_2$.  Further substitution of the complete power 
series of (\ref{pan1}) and (\ref{pan2}) into Eq.~(\ref{honse2}) shows 
that there are a sufficient number of arbitrary functions if and only 
if $\gamma_1=3$ and $\gamma_2=-3/2$.  This corresponds to the case 
studied in Ref.~\onlinecite{Sa91}.  Thus, the parameters that insure 
the Painlev\'{e} feature are a proper subset of the parameters that 
insure existence of an $N$-soliton solution of the HONSE.

To summarize, we have derived a $1$-soliton solution of the general 
HONSE, without any constraints on its coefficients except for the weak
inequality constraint $3\gamma_1+2\gamma_2 > 0$.   The $1$-soliton
width and intensity are related by the expression $I_s \tau_s^2 =
6/(3\gamma_1+2\gamma_2)$, so the intensity decreases with increasing
$(3\gamma_1+2\gamma_2)$ for a given $\tau_s$. We have shown that
$N$-soliton solutions exist when $\gamma_1=3$ and $\gamma_2$ is
arbitrary, thus extending significantly the results of
Ref.~\onlinecite{Hirota,Sa91}. The Painlev\'{e} plausibility condition
for integrability, $\gamma_1=-2\gamma_2=3$, has been shown to
be a proper subset of the conditions for $N$-soliton solutions, and
consistent with the integrable case found in Ref.~\onlinecite{Sa91},
while integrability of our more general case is still unknown.  Our
analysis does not exclude existence of $N$-soliton solutions nor does
it disprove integrability for arbitrary $\gamma_1$ and $\gamma_2$.

\acknowledgments
This work was supported in part by a grant from the US-Israel Binational
Science Foundation.


\begin{references}
\bibitem{Kodama} Y. Kodama and A. Hasegawa, IEEE J. Quantum Electron.
{\bf QE-23}, 5610 (1987).
\bibitem{Mitschke} F. M. Mitschke and L. F. Mollenauer, Opt. Lett. {\bf
11}, 657 (1986).
\bibitem{Hasegawa} A. Hasegawa, {\it Optical Solitons in Fibers}, (Springer,
Heidelberg, 1989).
\bibitem{Agrawal} G. P. Agrawal, {\it Nonlinear Fiber Optics}, (Academic 
Press, NY, 1989).
\bibitem{AS81}
M.A. Ablowitz and H. Segur, {\it Solitons and the inverse scattering 
transform}, (SIAM, Philadelphia, 1981).
\bibitem{Hasegawa2} A. Hasegawa and F. Tappert, Appl. Phys. Lett. {\bf
23}, 142 (1973).
\bibitem{Porsezian} K. Porsezian and K. Nakkeeran, Phys. Rev. Lett. {\bf
76}, 3955 (1996).
\bibitem{Weiss} J. Weiss, M. Tabor, and G. Carnevale, J. Math. Phys. (NY) 
{\bf 24}, 522 (1983).
\bibitem{Musette} M. Musette and R. Conte, J. Math. Phys. (NY) {\bf 32}, 1450
(1991).
\bibitem{Ablowitz} M. J. Ablowitz and
 P.A. Clarkson, {\it Solitons, Nonlinear 
Evolution Equations and Inverse Scattering}, (Cambridge University Press, 
Cambridge, 1991).
\bibitem{Hirota} R. Hirota, Phys. Rev. Lett. {\bf 27}, 1192 (1971); 
J. Math. Phys. (NY) {\bf 14}, 805 (1973).
\bibitem{Ba70}
 {\it The Pad\'{e} approximant in theoretical physics}, Eds.
  G.A. Baker and J.L. Gammel, (Academic Press, NY, 1970).
\bibitem{LW94}
Shan-liang Liu and Wen-zheng Wang, Phys. Rev. E {\bf 49}, 5726 (1994).
\bibitem{Sa91}
N. Sasa and J. Satsuma, J. Phys. Soc. Jpn. {\bf 60}, 409 (1991).
\bibitem{Ne85}
A.C. Newell, {\it Solitons in mathematics and physics}, (SIAM, 
Philadelphia, 1985).
 \bibitem{Ra89} 
A. Ramani, B. Grammaticos, and T. Bountis, Phys. Rep. {\bf 180}, 159 (1989).
\end{references}
\end{document}